# Application of a variational hybrid quantum-classical algorithm to heat conduction equation

Y. Y. Liu[a], Z. Chen[b], C. Shu[a,*], S. C. Chew[c], B. C. Khoo[a] and X. Zhao[d]

[a] Department of Mechanical Engineering, National University of Singapore, 10 Kent Ridge Crescent, Singapore 119260

[b] School of Naval Architecture, Ocean and Civil Engineering, Shanghai Jiao Tong University, Shanghai, China 200240

[c] National University of Singapore, Singapore 117411, Republic of Singapore

[d] Institute of High Performance Computing, Agency for Science, Technology and Research, 1 Fusionopolis Way, Singapore 138632, Singapore

## Abstract

The prosperous development of both hardware and algorithms for quantum computing (QC) potentially prompts a paradigm shift in scientific computing in various fields. As an increasingly active topic in QC, the variational quantum algorithm (VQA) leads a promising tool for solving partial differential equations on Noisy Intermediate Scale Quantum (NISQ) devices. Although a clear perspective on the advantages of QC over classical computing techniques for specific mathematical and physical problems exists, applications of QC in computational fluid dynamics to solve practical flow problems, though promising, are still in an early stage of development. To explore QC in practical simulation of flow problems, this work applies a variational hybrid quantum-classical algorithm, namely the variational quantum linear solver (VQLS), to resolve the heat conduction equation through finite difference discretization of the Laplacian operator. Details of VQLS implementation are discussed by various test instances of linear systems. Finally, the successful statevector simulations of the heat conduction equation in one and two dimensions demonstrate the validity of the present algorithm by proof-of-concept results. In addition, the heuristic scaling for the heat conduction problem indicates that the time complexity of the present approach is logarithmically dependent on the precision $\varepsilon$ and linearly dependent on the number of qubits $n$.

* Corresponding author, E-mail: mpeshuc@nus.edu.sg (C. Shu).

## 1. Introduction

The slower pace of benefits from Moore's law has prompted many discussions about its validity over the recent years [1]. Limited by the increasingly smaller sizes of silicon processes, groundbreaking advance of computational performance within the classical computing framework is becoming more and more effortful. Different from the conventional computing paradigm, quantum computing (QC) performs the calculations through the physical manipulation of quantum systems, which is a potential game-changer in the scientific computing community and may unleash unprecedented growth in computational capabilities.

Recent decades have witnessed the rapid development of algorithms and hardware for QC technology. A quantum advantage has been demonstrated for the first time in a benchmark task of sampling the output of a pseudo-random quantum circuit [2]. Such a milestone inspires numerous explorations of QC in various scientific and engineering areas, such as material science, chemistry, machine learning and fluid dynamics, especially in current Noisy Intermediate Scale Quantum (NISQ) [3] era. In particular, the computational fluid dynamics (CFD), which is closely associated with the scientific computing techniques, naturally benefits from the promising computation power of QC [4,5]. The essence of CFD is to discretize the partial differential equations (PDEs) describing fluid phenomena into a set of algebraic equations which can be numerically resolved on computational platforms. Applications of CFD to engineering problems place high demands in computational scale and efficiency, many of which are still beyond current computational capabilities. Thus, for sophisticated CFD missions, QC indeed provides a potential alternative to classical computing framework [6-9].

Within the context of quantum computing for CFD (QCFD), many quantum algorithms have been developed. In the literature, typical approaches with practical implementations on the quantum simulators or real devices include quantum machine learning (QML) [10-12], Harrow-Hassidim-Lloyd (HHL) algorithm [13-17] and variational quantum algorithms [18-20].

Among them, the HHL algorithm was initially proposed as a heuristic algorithm to solve the linear problem of $A\mathbf{x} = \mathbf{b}$ where $A$ denotes a Hermitian sparse $N \times N$ matrix and $\mathbf{b}$ is a unit vector. This work [13] reported an exponential speedup for the HHL algorithm. Subsequently, many research studies applied it as a subroutine to solve PDEs [14-17,21,22]. However, it remains unclear whether the utilization of the HHL algorithm can achieve the exponential speedup for practical fluid problems involving a much larger size and more complex structure of the matrix $A$. At present, it has been demonstrated on the circuit quantum computers that the HHL-type linear solvers are limited to $4 \times 4$ matrices [23,24]. Apart from the limited computational scale, in the NISQ era, short coherence time of noisy quantum devices which are not fault-tolerant and cannot perfectly control the qubits could be the major obstacle to the application of the HHL algorithm.

Unlike the HHL-type algorithms, the VQA algorithms are more adaptive to the NISQ hardware and therefore are more realistic strategies to execute the actual quantum advantage. In practice, the successful application of VQA algorithms to solve linear systems refers to the variational quantum linear solver (VQLS) [25-28]. The VQLS algorithm can also serve as a subroutine similar to the HHL algorithm. Specifically, the system of linear equations obtained by discretizing PDEs with proper numerical schemes like the finite difference method (FDM) [31,32] and finite element method (FEM) [33] is solved by the VQLS algorithm. Such a technique has been applied to solve the Poisson equation [20,34,35]. However, practical applications of the VQLS algorithm to flow problems of engineering interest remain scarce. Further explorations are needed to quantify the actual quantum advantage of a heuristic quantum technology over the classical numerical methods.

Here, we demonstrate a practical utility of the VQLS algorithm by solving the heat conduction problem governed by Laplace's equation. The Laplace's equation is a special case of the Poisson's equation and has been broadly used in fluid dynamics. Although there exist several studies of such heat transfer applications using other non-variational quantum algorithms [15,36,37], in the NISQ era, those methods may be less competitive than the incorporation with the VQLS algorithm which employs a shallower-depth quantum circuit for efficient evaluation of a cost function. Inspired by the aforementioned works [20,34,35] about the Poisson's equation and based on the fact of relatively easy implementations for the Laplace's equation, we can focus on assessing the performance of the VQLS algorithm in this practical problem. This work may lay a foundation for further applications of the VQLS algorithm to more complicated fluid flow problems.

The remainder of this paper is organized as follows. Section 2 states the heat conduction equation and FDM algorithm for its discretization. The VQLS algorithm is then introduced. In Section 3, technical details of the VQLS implementation are discussed, followed by its application to solve the heat conduction problem in Section 4. Section 5 concludes this paper with some remarks.

## 2. Methodology

### 2.1. Governing equations and finite difference discretization

In this paper, the steady state heat transfer problem which can be described by a system of linear PDEs is considered via the following $d$-dimensional Laplace's equation without a source term:

$$\nabla^2 T = \frac{\partial^2 T}{\partial x_1^2} + \cdots \frac{\partial^2 T}{\partial x_d^2} = 0, \tag{1}$$

where $T$ denotes temperature. Dirichlet boundary conditions are enforced with the temperature $T_1$ for the bottom boundary and $T_2$ for the upper boundary. The other boundaries are periodic. After discretizing Eq. (1) by the FDM, a linear system can be obtained. Specifically, in two dimensions, we have the following expression.

$$\nabla^2 T = \frac{\partial^2 T}{\partial x^2} + \frac{\partial^2 T}{\partial y^2} = 0. \tag{2}$$

Discretizing Eq. (2) with the central difference yields

$$\frac{T_{i+1,j} - 2T_{i,j} + T_{i-1,j}}{\Delta x^2} + \frac{T_{i,j+1} - 2T_{i,j} + T_{i,j-1}}{\Delta y^2} = 0. \tag{3}$$

When the uniform mesh is used, i.e., the mesh spacings $\Delta x = \Delta y$, the following equation can be further obtained.

$$T_{i+1,j} + T_{i-1,j} + T_{i,j+1} + T_{i,j-1} - 4T_{i,j} = 0. \tag{4}$$

The subscripts $i$ and $j$ in above equations indicate the grid indices in the $x$- and $y$- axes, respectively. Considering that there are $N+2$ grid points in both directions, $i$ and $j$ can be 1, 2, …, $N+2$. For illustration purposes, Fig. 1 presents a schematic diagram of the heat conduction problem on the uniform mesh.

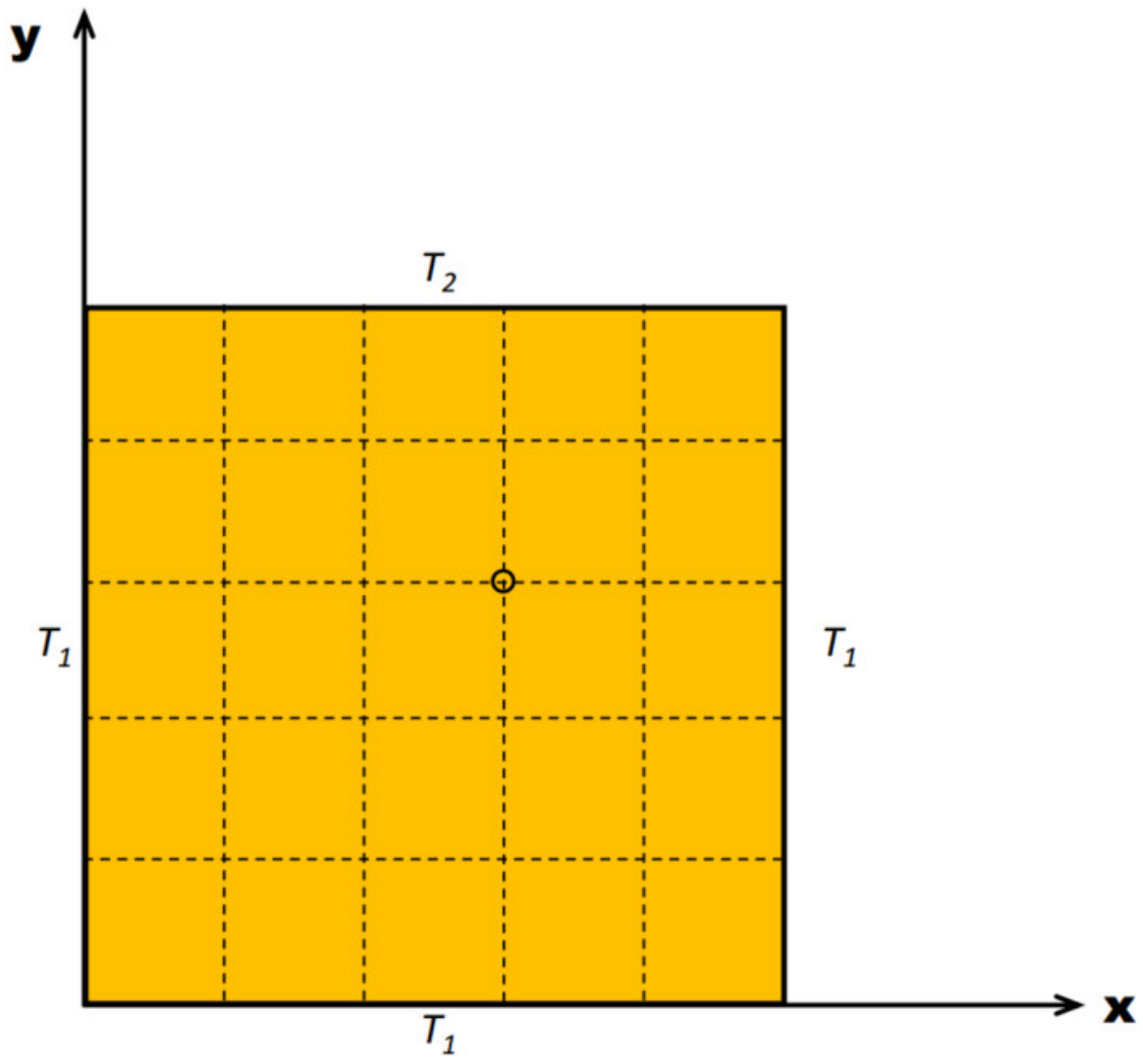

Fig. 1. Schematic diagram of heat conduction problem.

Given the Dirichlet boundary conditions and Eq. (3) for each interior grid point in the whole flow domain, such a problem can be numerically resolved via acquiring solutions to the following system of coupled linear equations

$$A\mathbf{x} = \mathbf{b}, \tag{5}$$

where

$$A = \begin{bmatrix} -4 & 1 & 0 & & & 0 & 1 & 0 & & & 0 \\ 1 & -4 & \ddots & \ddots & & & \ddots & \ddots & \ddots & & \\ 0 & \ddots & \ddots & \ddots & \ddots & & & \ddots & \ddots & \ddots & \\ & \ddots & \ddots & \ddots & 1 & \ddots & & & \ddots & \ddots & 0 \\ & & \ddots & 1 & \ddots & 0 & \ddots & & & \ddots & 1 \\ 0 & & & \ddots & 0 & \ddots & 1 & \ddots & & & 0 \\ 1 & \ddots & & & \ddots & 1 & \ddots & \ddots & \ddots & & \\ 0 & \ddots & \ddots & & & \ddots & \ddots & \ddots & \ddots & \ddots & \\ & \ddots & \ddots & \ddots & & & \ddots & \ddots & \ddots & 1 & 0 \\ & & \ddots & \ddots & \ddots & & & \ddots & 1 & -4 & 1 \\ 0 & & & 0 & 1 & 0 & & & 0 & 1 & -4 \end{bmatrix}_{N^2 \times N^2} \tag{6}$$

$$\mathbf{x} = \begin{bmatrix} T_{11} & T_{21} & \dots & T_{N1} & T_{12} & T_{22} & \dots & T_{N2} & \dots & T_{N-1N} & T_{NN} \end{bmatrix}^T_{N^2}$$

$$\mathbf{b} = \begin{bmatrix} 0 & 0 & \dots & \dots & \dots & 0 & -1 & \dots & \dots & -1 & -1 \end{bmatrix}^T_{N^2}.$$

Clearly, $A$ is a $N^2 \times N^2$ sparse matrix and $\mathbf{b}$ denotes the known vector by setting $T_1$

= 0 and $T_2$ = 1. Note that the number of interior grid points along each direction is $N = 2^{n/2}$ where $n$ represents the number of qubits used.

Through similar process, the linear system for the one-dimensional case can be derived as follows.

$$A = \begin{bmatrix} -2 & 1 & & & & & \\ 1 & & & & & & \\ & & \ddots & \ddots & & & \\ & & \ddots & & & & \\ & & & & \ddots & \ddots & \\ & & & & \ddots & & 1 \\ & & & & & 1 & -2 \end{bmatrix}_{N \times N} \tag{7}$$

$$\mathbf{b} = \begin{bmatrix} 0 & 0 & 0 & \ldots & 0 & 0 & 0 & -1 \end{bmatrix}^T_N .$$

In this case, the number of interior grid points is $N = 2^n$ with the number of qubits $n$.

Solutions of the above linear systems can be solved by the VQLS algorithm introduced in Subsection 2.2, which directly resolves the heat conduction problem without any iteration.

## 2.2. Variational quantum linear solver to solve linear systems of equations

The VQLS is a variational quantum algorithm for solving linear systems of equations on NISQ quantum computers and possesses superior efficiency over classical computational methods. To be specific, for the linear systems in Eq. (5), the VQLS can find a normalized $|\mathrm{x}\rangle$ to fulfill the relationship $A|\mathrm{x}\rangle = |\mathrm{b}\rangle$ where $|\mathrm{b}\rangle$ denotes the quantum state prepared from the known vector **b**. Fig. 2 depicts an overview of the VQLS algorithm. The inputs to this algorithm are the matrix $A$ given as a linear combination of unitary matrices $A_m$ with the coefficients $c_m$ and a short-depth quantum circuit $U$ such that $|\mathrm{b}\rangle = U|0\rangle$. This process can be completed through state preparation. Then the cost function $C(\boldsymbol{\alpha})$ is constructed and evaluated with a devised parameterized ansatz $V(\boldsymbol{\alpha})$. Through the hybrid quantum-classical optimization loop, the optimal parameters $\boldsymbol{\alpha}^*$ for the ansatz circuit can be found when the cost function $C(\boldsymbol{\alpha})$ achieves the convergence criterion. At the end of the feedback loop, the ansatz $V(\boldsymbol{\alpha}^*)$ prepares the quantum state $|\mathrm{x}^*\rangle$ that is proportional to the solution $\mathbf{x}$, as the final output. Here, for clarity, only three processes, namely, state preparation, ansatz and cost function are illustrated. Interesting readers may refer to Ref. [25] for more details of the VQLS algorithm.

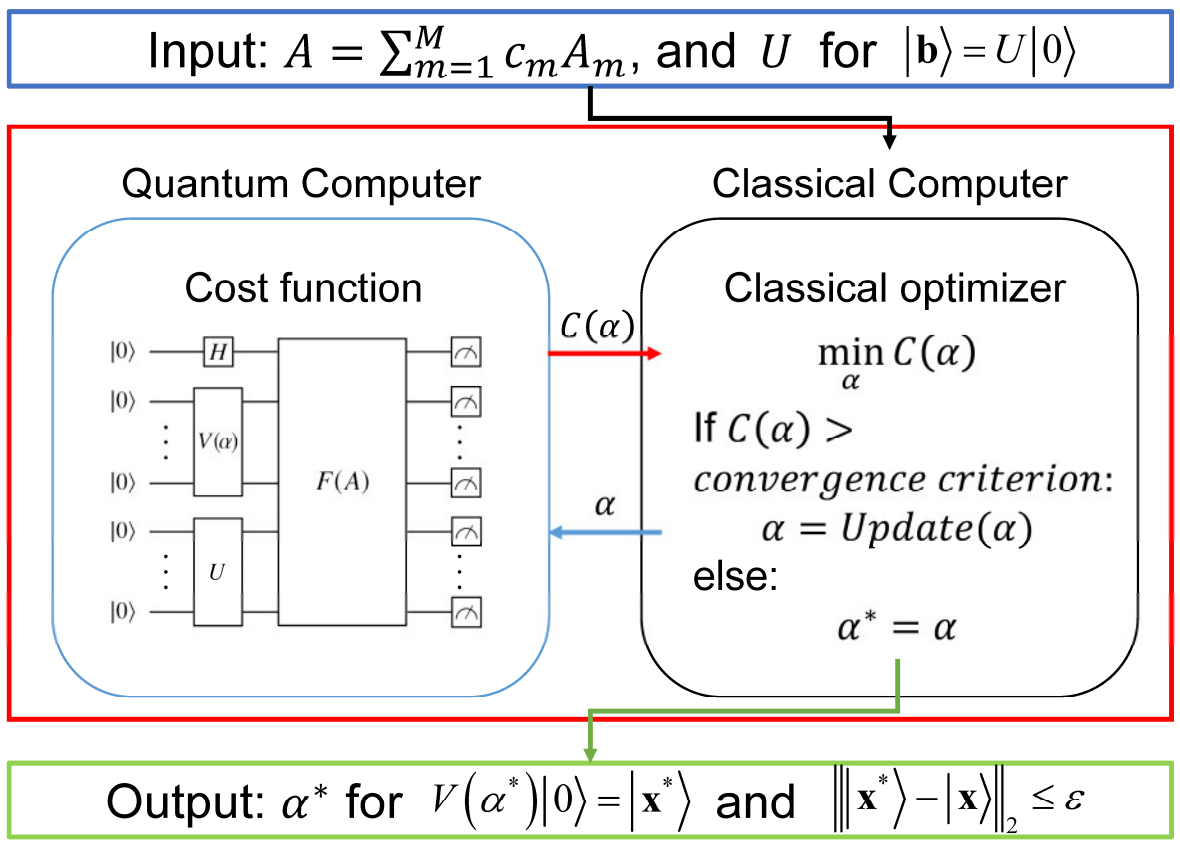

Fig. 2. Basic VQLS algorithm schematic diagram.

## A. State preparation

The input to the VQLS algorithm requires the quantum states representing the matrix $A$ and the vector $\mathbf{b}$. In practical implementation, the matrix $A$ is decomposed into a linear combination of $M$ unitary matrices $A_m$ with their complex coefficients $c_m$ as follows:

$$A = \sum_{m=1}^{M} c_m A_m. \tag{8}$$

For the decomposition, one popular approach [20,25,27,34] is based on the identity $I$ and the following Pauli gates basis of $X$, $Y$ and $Z$.

$$X = \begin{bmatrix} 0 & 1 \\ 1 & 0 \end{bmatrix}, Y = \begin{bmatrix} 0 & -i \\ i & 0 \end{bmatrix}, Z = \begin{bmatrix} 1 & 0 \\ 0 & -1 \end{bmatrix}. \tag{9}$$

For example, for a matrix $A$ with the size $2^n \times 2^n$, the Pauli basis can be selected as $\sigma_n = \{P_1 \otimes \cdots \otimes P_l\}^n$, $P_l \in \{I, X, Y, Z\}$. And $A_m$ belongs to $\sigma_n$. Its corresponding coefficients $c_m$ are determined via $c_m = \mathrm{Tr}(A_m \cdot A)/2^n$ where Tr represents the trace.

A normalized complex vector of a quantum state $|\mathrm{b}\rangle$ should also be prepared. This can be fulfilled by applying a unitary operation $U$ to the ground state $|0\rangle$, i.e.,

$$|\mathrm{b}\rangle = U|0\rangle. \tag{10}$$

The unitary operation $U$ may be found by using the method proposed by Shende et al. [30].

**B. Hardware-efficient ansatz**

The VQLS employs an ansatz for the gate sequence $V(\boldsymbol{\alpha})$ which simulates a potential solution $|\mathrm{x}\rangle = V(\boldsymbol{\alpha})|0\rangle$. In fact, it is free to choose the ansatz for specific problems, while we will follow the popular choice of using the fixed structure hardware-efficient ansatz [25,29]. From the example shown in Fig. 3, such an ansatz consists of multiple layers (each layer is identified in the green square) of controlled-$Z$

gates across alternating pairs of neighboring qubits entangled by $R_y(\alpha)$ rotation gates. In the example in Fig. 3, all parameters $\alpha$ are set to 1 for illustration purposes. Noteworthily, during every run of the quantum circuit, the structure of quantum gates remains the same and only the parameters $\alpha$ in $R_y$ gates change.

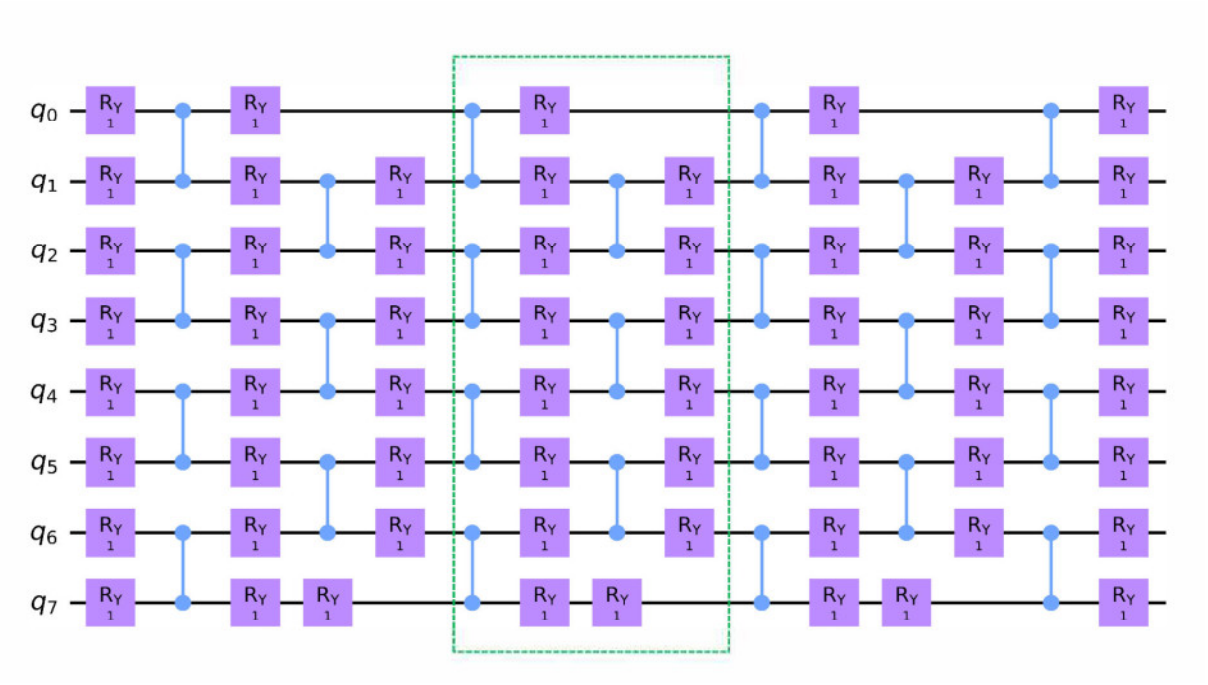

Fig. 3. An example of the fixed structure hardware-efficient ansatz.

## C. Cost function

With the ansatz $V(\boldsymbol{\alpha})$, the state $\left|\mathrm{x}(\boldsymbol{\alpha})\right\rangle = V(\boldsymbol{\alpha})\left|0\right\rangle$ can be prepared. Hereinafter, $\left|\mathrm{x}(\boldsymbol{\alpha})\right\rangle$ is denoted as "$\left|\mathrm{x}\right\rangle$" for simplicity. Since the VQLS algorithm aims to minimize the cost function $C(\boldsymbol{\alpha})$, when the state $\left|\Phi\right\rangle = A\left|\mathrm{x}\right\rangle$ is nearly proportional to $|\mathrm{b}\rangle$, the value of the cost function should be very small. Conversely, the cost function will be very large when $\left|\Phi\right\rangle$ is close orthogonal to $|\mathrm{b}\rangle$. Thus, the following cost function is introduced

$$\begin{aligned} C_{Gp}(\boldsymbol{\alpha}) &= C_{Gp} \\ &= \mathrm{Tr}\left(A\left|\mathrm{x}\right\rangle\left\langle\mathrm{x}\right|A^{\dagger}\left(I - \left|\mathrm{b}\right\rangle\left\langle\mathrm{b}\right|\right)\right) \\ &= \mathrm{Tr}\left(\left\langle\mathrm{x}\right|A^{\dagger}\left(I - \left|\mathrm{b}\right\rangle\left\langle\mathrm{b}\right|\right)A\left|\mathrm{x}\right\rangle\right) \\ &= \left\langle\mathrm{x}\right|H_P\left|\mathrm{x}\right\rangle, \end{aligned} \tag{11}$$

with the Hamiltonian $H_P$ defined as

$$H_P = A^{\dagger}\left(I - |\mathrm{b}\rangle\langle\mathrm{b}|\right)A. \tag{12}$$

Generally, normalizing the cost function is necessary to increase the accuracy of the algorithm. In practice, the cost function $C_{Gp}$ should be divided by the norm of $|\Phi\rangle$, which yields

$$C_p = \frac{\langle\Phi|\left(I - |\mathrm{b}\rangle\langle\mathrm{b}|\right)|\Phi\rangle}{\langle\Phi|\Phi\rangle} = \frac{\langle\Phi|\Phi\rangle - \langle\Phi|\mathrm{b}\rangle\langle\mathrm{b}|\Phi\rangle}{\langle\Phi|\Phi\rangle} = 1 - \frac{|\langle\mathrm{b}|\Phi\rangle|^2}{\langle\Phi|\Phi\rangle}. \tag{13}$$

The minimization of $C_p$ with respect to the variational parameters leads towards the problem solution.

Clearly, there are two values to be calculated to evaluate the cost function $C_p$, namely $\langle\Phi|\Phi\rangle$ and $|\langle\mathrm{b}|\Phi\rangle|^2$. Given the linear combination of unitary matrices for the matrix $A$ and the ansatz $V(\boldsymbol{\alpha})$, $\langle\Phi|\Phi\rangle$ can be computed by

$$\begin{aligned}\langle\Phi|\Phi\rangle &= \left(A|\mathrm{x}\rangle\right)^{\dagger} A|\mathrm{x}\rangle \\ &= \sum_{m=1}^{M}\sum_{m'=1}^{M}\langle 0|V^{\dagger}A_{m'}^{\dagger}A_m V|0\rangle c_m c_{m'}^{*},\end{aligned} \tag{14}$$

and $|\langle\mathrm{b}|\Phi\rangle|^2$ can be gotten via

$$\begin{aligned}|\langle\mathrm{b}|\Phi\rangle|^2 &= |\langle\mathrm{b}|A|\mathrm{x}\rangle|^2 \\ &= \sum_{m=1}^{M}\sum_{m'=1}^{M}\langle 0|U^{\dagger}A_m V|0\rangle\langle 0|V^{\dagger}A_{m'}^{\dagger}U|0\rangle c_m c_{m'}^{*}.\end{aligned} \tag{15}$$

All expectation values of these two items can be estimated by a standard quantum computation technique called the Hadamard Test. In quantum computation, the Hadamard test is a method utilized to construct a random variable whose expectation value is the expected real part of the observed value of a quantum state $|\varphi\rangle$, i.e., $\mathrm{Re}\langle\varphi|U_H|\varphi\rangle$, with respect to a unitary operator $U_H$. The technical details of the Hadamard test can be found in Refs. [25] and [34].

However, such a global expression of the cost function in Eq. (13) requires that all unitary operations ($U$, $U^\dagger$, $A^\dagger_m$, $A_m$, $V^\dagger$ and $V$) are controlled by an external ancillary qubit. This might be experimentally challenging in the Hadamard test, especially when the ansatz $V$ consists of many layers. Thus, in practical implementation, a local version of the cost function which is easier to measure and leads towards the same optimal solution is utilized. This local cost function replaces the projector |0⟩⟨0| in Eq. (15) with the following positive operator:

$$P = \frac{1}{2} I + \frac{1}{2n} \sum_{l=0}^{n-1} Z_l, \tag{16}$$

where $Z_l$ denotes the Pauli $Z$ operator locally acted on the $l$th qubit. In this way, we can get

$$\left|\langle \mathrm{b}|\Phi\rangle\right|^2 = \sum_{m=1}^{M} \sum_{m'=1}^{M} \langle 0| V^\dagger A_{m'}^\dagger U P U^\dagger A_m V |0\rangle c_m c_{m'}^*. \tag{17}$$

Then the local cost function $C$ will be

$$C = 1 - \frac{\sum_{m=1}^{M} \sum_{m'=1}^{M} \langle 0| V^\dagger A_{m'}^\dagger U P U^\dagger A_m V |0\rangle c_m c_{m'}^*}{\sum_{m=1}^{M} \sum_{m'=1}^{M} \langle 0| V^\dagger A_{m'}^\dagger A_m V |0\rangle c_m c_{m'}^*}. \tag{18}$$

Substituting the definition of $P$ into the expression yields

$$C = \frac{1}{2} - \frac{1}{2n} \frac{\sum_{l=0}^{n-1} \sum_{m=1}^{M} \sum_{m'=1}^{M} u_{m,m',l} c_m c_{m'}^*}{\sum_{m=1}^{M} \sum_{m'=1}^{M} u_{m,m',-1} c_m c_{m'}^*}, \tag{19}$$

with the coefficients

$$u_{m,m',l} = \langle 0 | V^{\dagger} A_{m'}^{\dagger} U Z_l U^{\dagger} A_m V | 0 \rangle, \tag{20}$$

where $Z_l$ is replaced with the identity if $l = -1$, i.e.,

$$u_{m,m',-1} = \langle 0 | V^{\dagger} A_{m'}^{\dagger} A_m V | 0 \rangle. \tag{21}$$

The complex coefficients $u_{m,m',l}$ can be experimentally estimated with the Hadamard test where only the unitaries $A^{\dagger}_m$, $A_m$ and $Z_l$ need to be applied in a controlled way. Finally, the problem can be solved by minimizing the local cost function $C$.

## 3. Implementation Tests of Variational Quantum Linear Solver

Based on the theoretical background of the VQLS algorithm illustrated in Subsection 2.2, this section applies this algorithm to solve a series of test problems on the quantum simulator as a proof-of-concept. Generally, there are three factors affecting the convergence [34], i.e., the number of shots which determines numerical noise on evaluation process, the ansatz which is associated with the number of qubits and the classical optimizer which is related to the number of variables. In this work, we employ a hardware-efficient ansatz $V(\boldsymbol{\alpha})$ as described in Subsection 2.2. For the selection of the classical optimizer, the comparison of various optimizers reported in Ref. [39] offers important guidance. Here, the gradient-descent optimizer with momentum [40,41] is utilized. Since the initialization parameters of the ansatz can greatly affect the optimization process, they are set randomly to ensure generality.

Issues concerning the finite sampling of the quantum circuit output (number of shots) and time complexity (heuristic scaling) for the VQLS algorithm are discussed with the simulated results. Although these discussions basically are problem-specific, the analyses are deemed relevant and indeed provide proper guidance for other applications. All quantum simulations are implemented using the Xanadu's Pennylane open-source library [38] with a statevector simulator as a backend. In the comparison hereinafter, the results obtained by the classical solver and VQLS solver are noted as "classic" and "VQLS", respectively.

### 3.1 Definition of the test case

Here, the illustrative test instance is given by a matrix $A$ with the size of $N \times N$ and a state vector $|\mathrm{b}\rangle$ as follows.

$$\begin{aligned} A &= c_0 A_0 + c_1 A_1 + c_2 A_2 \\ &= c_0 I^{\otimes n} + 0.2 X_0 Z_1 + 0.2 X_0, \\ |\mathrm{b}\rangle &= U|0\rangle = \prod_{l=0}^{n} H_l |0\rangle. \end{aligned} \tag{22}$$

where $n$ represents the number of qubits used and $N = 2^n$. $H$ is the Hadamard gate which is represented by the following Hadamard matrix.

$$H = \frac{1}{\sqrt{2}} \begin{bmatrix} 1 & 1 \\ 1 & -1 \end{bmatrix}. \tag{23}$$

The coefficients $c_0$ , $c_1$ and $c_2$ in Eq. (22) determine the matrix property. In this test, $c_1$ and $c_2$ are fixed at $c_1 = c_2 = 0.2$, while $c_0$ is adjusted to change the condition number $k$ for the matrix $A$. Accordingly, the matrix $A$ has the following expression.

$$
A = \begin{array}{r} \\ \\ \\ \\ \\ \text{row:}N/2+1 \\ \\ \text{row:}3N/4 \\ \\ \\ \\ \end{array}
\begin{array}{c}
\begin{array}{cccccccccc} & & & & \text{col:}N/2+1 & & \text{col:}3N/4 & & & \end{array} \\
\left[\begin{array}{cccccccccc}
c_0 & & & & 0.4 & & & & & \\
 & \ddots & & & & \ddots & & & & \\
 & & & & & & 0.4 & & & \\
 & & & & & & & 0 & & \\
0.4 & & & & \ddots & & & & \ddots & \\
 & \ddots & & & & \ddots & & & & 0 \\
 & & 0.4 & & & & & & & \\
 & & & 0 & & & & & & \\
 & & & & \ddots & & & & \ddots & \\
 & & & & & 0 & & & & c_0
\end{array}\right]_{N\times N}
\end{array} . \tag{24}
$$

Clearly, the matrix $A$ is sparse. The condition number is fixed at 2.3333 when $c_0 = 1$ no matter how many qubits are used.

### 3.2 Results and discussion

First, the test instance of matrix size 1024 × 1024 with $c_0 = 1$, i.e., the number of qubits $n = 10$, is solved to validate the VQLS algorithm for large matrix problems. Fig. 4 plots the obtained solutions, in which the $x$-axis denotes the index of the solution point and the $y$-axis represents the value of the solution. Note that the perfect simulation with the statevector simulator outputs the analytical state of the solutions without any noise. These quantum solutions are extracted when the local cost function reaches the convergence threshold of $10^{-9}$. Good agreement between the results computed by the VQLS and classical solvers are achieved, as shown in Fig. 4, which indicates the validity of the VQLS algorithm in solving a large linear system.

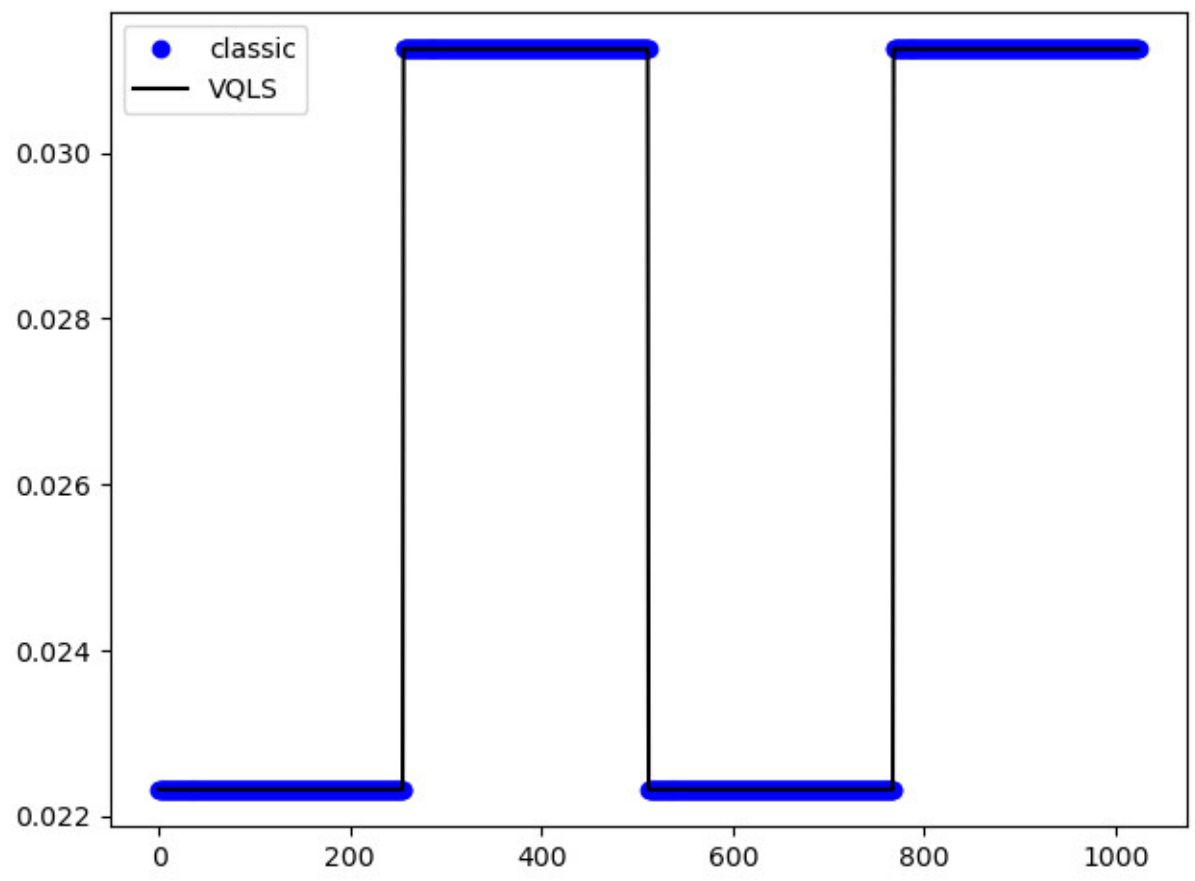

Fig. 4. Analytical state obtained by using the VQLS algorithm with $n = 10$.

Although the statevector simulator utilized here conducts the perfect simulations and the analytical quantum state is obtained, the result of a quantum circuit is inherently probabilistic. The practical accuracy would highly depend on the sampling of the circuit outcomes. In other words, the quantum device would have to sample the circuit many times (number of shots) to generate the probabilities over the basis states instead of calculating the probabilities numerically. Thus, the effect of the noise or errors resulted from the number of shots is worth investigation and would be analyzed here.

In the simulation, four linear systems given in Eq. (22) with $c_0 = 1$ of the matrix size $8 \times 8$ ($n = 3$), $16 \times 16$ ($n = 4$), $32 \times 32$ ($n = 5$) and $256 \times 256$ ($n = 8$) are solved. As shown from Fig. 5 to Fig. 8, when the number of shots is selected as 3000, the quantum solutions after sampling differ from the analytical state. The same observations can be found from the comparison of the probability distributions. As the matrix size increases (from $8 \times 8$ to $256 \times 256$), these discrepancies become larger.

These phenomena are expected due to the increasing depth and complexity of the quantum circuit for solving larger linear systems.

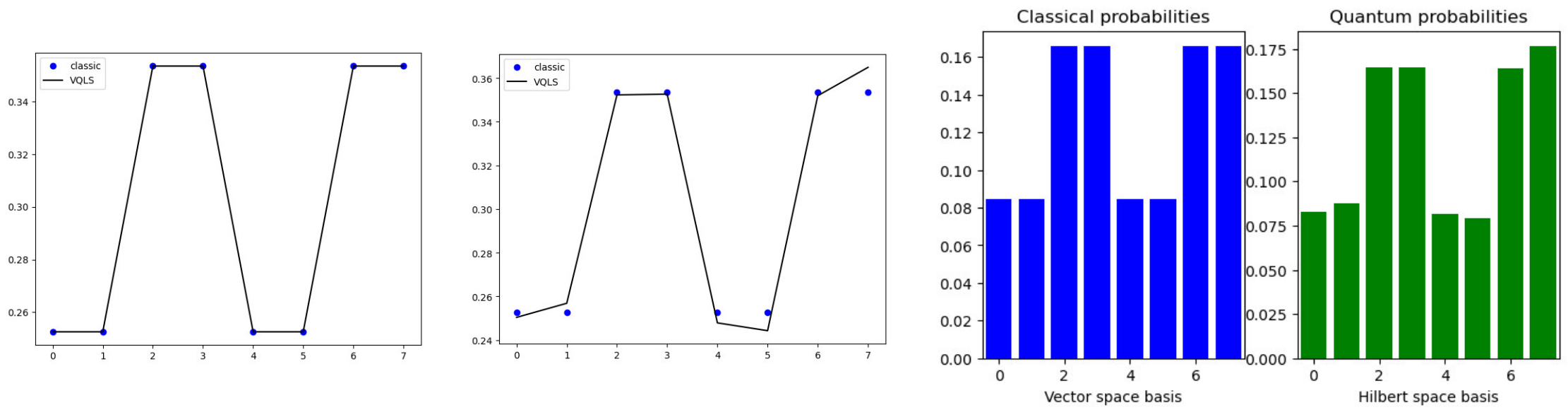

(a) Analytical state

(b) Quantum solutions with 3000 shots

(c) Quantum state probabilities with 3000 shots

Fig. 5. Results obtained by using the VQLS algorithm with $n = 3$: (a) Analytical state, (b) quantum solutions with 3000 shots and (c) quantum state probabilities with 3000 shots.

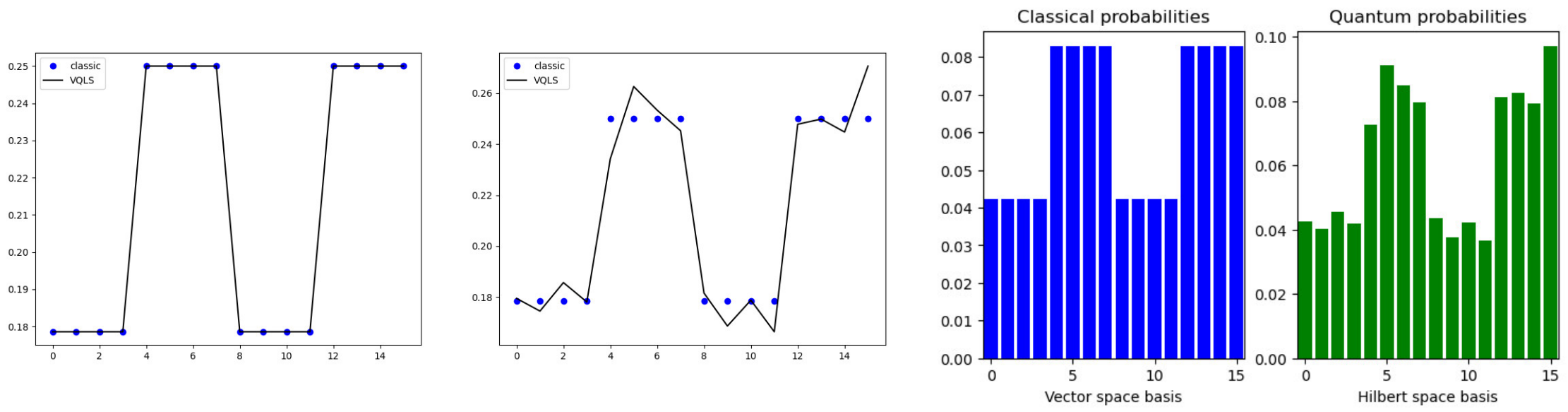

(a) Analytical state

(b) Quantum solutions with 3000 shots

(c) Quantum state probabilities with 3000 shots

Fig. 6. Results obtained by using the VQLS algorithm with $n = 4$: (a) Analytical state, (b) quantum solutions with 3000 shots and (c) quantum state probabilities with 3000 shots.

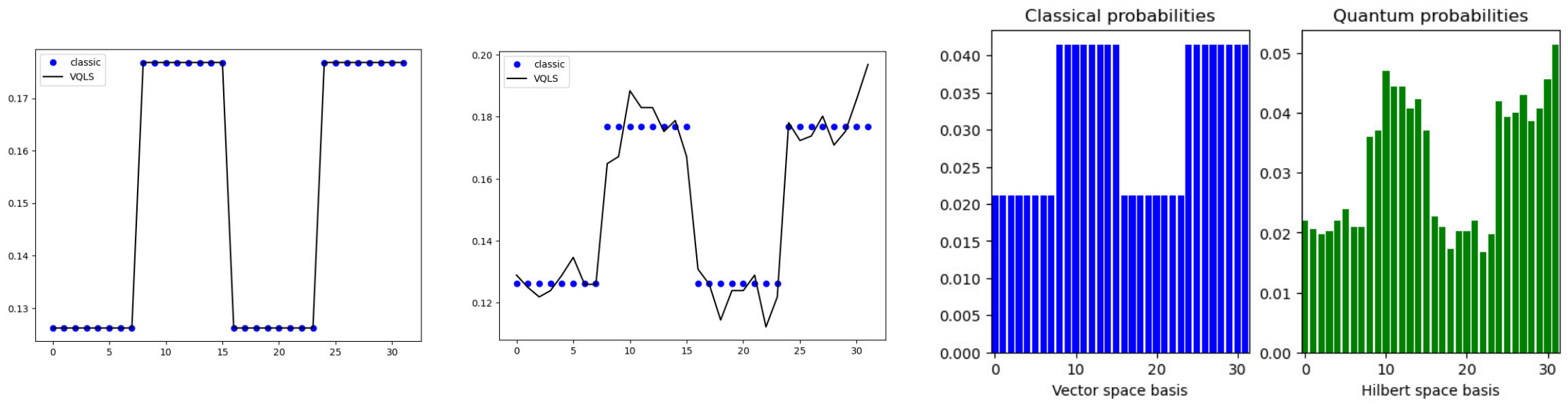

(a) Analytical state (b) Quantum solutions with 3000 shots (c) Quantum state probabilities with 3000 shots

Fig. 7. Results obtained by using the VQLS algorithm with $n$ = 5: (a) Analytical state, (b) quantum solutions with 3000 shots and (c) quantum state probabilities with 3000 shots.

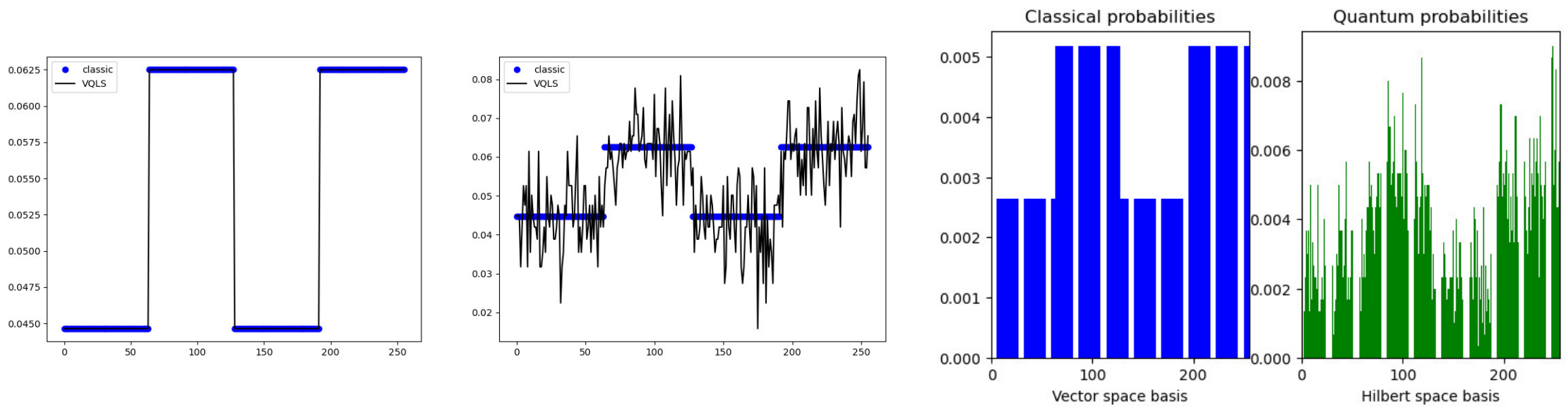

(a) Analytical state (b) Quantum solutions with 3000 shots (c) Quantum state probabilities with 3000 shots

Fig. 8. Results obtained by using the VQLS algorithm with $n$ = 8: (a) Analytical state, (b) quantum solutions with 3000 shots and (c) quantum state probabilities with 3000 shots.

Furthermore, the dependence of the precision $\varepsilon$ on the number of shots is heuristically evaluated. Fig. 9 plots the relationships between the number of shots and the precision. For each data point, we implemented and averaged over 10 executions of the VQLS solver. In the log-log scale, it is clear that the dependence on the number of shots for the precision appears to be linear. As the number of qubits becomes higher, more shots are required to guarantee the same precision. When the same number of shots is chosen, lower accuracy can be achieved for the cases using the larger number of qubits, which accords with the previous observation (in cases using 3000 shots). A reasonable hypothesis for these outcomes would be that the increase of the number of qubits results in more variables in the ansatz to be optimized and a deeper quantum circuit. Overall, the high fidelity demands a greater number of shots, potentially threatening the scaling of the VQLS algorithm.

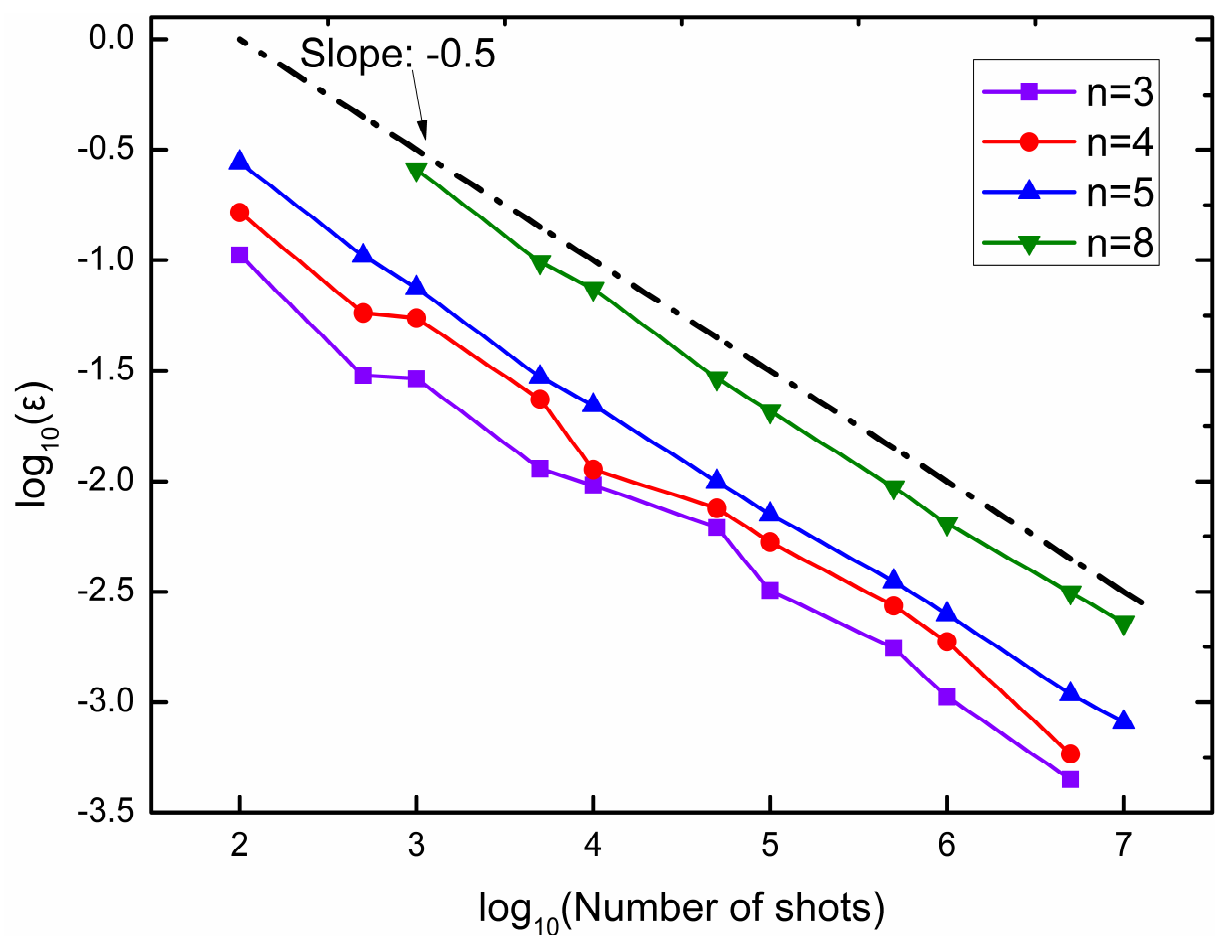

Fig. 9. Logarithmic plots of precision ($\varepsilon$) versus number of shots.

The time complexity of the VQLS is explored as well by studying its dependence on the precision $\varepsilon$, the number of qubits $n$ and the condition number $k$. Specifically, for the test instance given in Eq. (22), the scaling of the VQLS solver with $\varepsilon$ and $n$ would be heuristically assessed by fixing $c_0 = 1$, determining that $k$ remains to be 2.3333 while $n$ changes. To conduct the heuristic scaling with $k$, the coefficients $c_0 =$ 0.03, 0.05, 0.1, 1 and 2 for respectively $k$ = 14.3333, 9, 5, 2.3333 and 1.5 are used with $n$ = 4. Note that for the heuristic scaling, the perfect simulation is conducted without finite sampling. The run time of the solver is quantified with the evaluations-to-solution. This metric refers to the number of exact cost function evaluations during the optimization which can guarantee a desired precision $\varepsilon$. In all studies, over 10 executions of the VQLS solver are averaged to get the evaluations-to-solution.

In Fig. 10 where the $x$-axis is plotted in the logarithmic scale, the change of evaluations-to-solution with $1/\varepsilon$ is found to be linear for various $n$, implying that the dependence on $1/\varepsilon$ is logarithmic. The dependence on the number of qubits $n$ (or the matrix size $N$) is heuristically scaled in Fig. 11, where the precision $\varepsilon$ is specified as 0.001. The results evidently determine that the evaluations-to-solution is linearly dependent on $n$, which means the logarithmic dependence on $N$. Fig. 12 depicts the evaluations-to-solution versus $k$ when the particular precisions $\varepsilon$ = 0.01 and 0.03 are guaranteed. It is shown that as $k$ grows, the evaluations-to-solution exhibits a sub-linear increase. These observations are analogous to those in Ref. [25], which confirms the competitive efficiency of the VQLS algorithm compared with the classical methods.

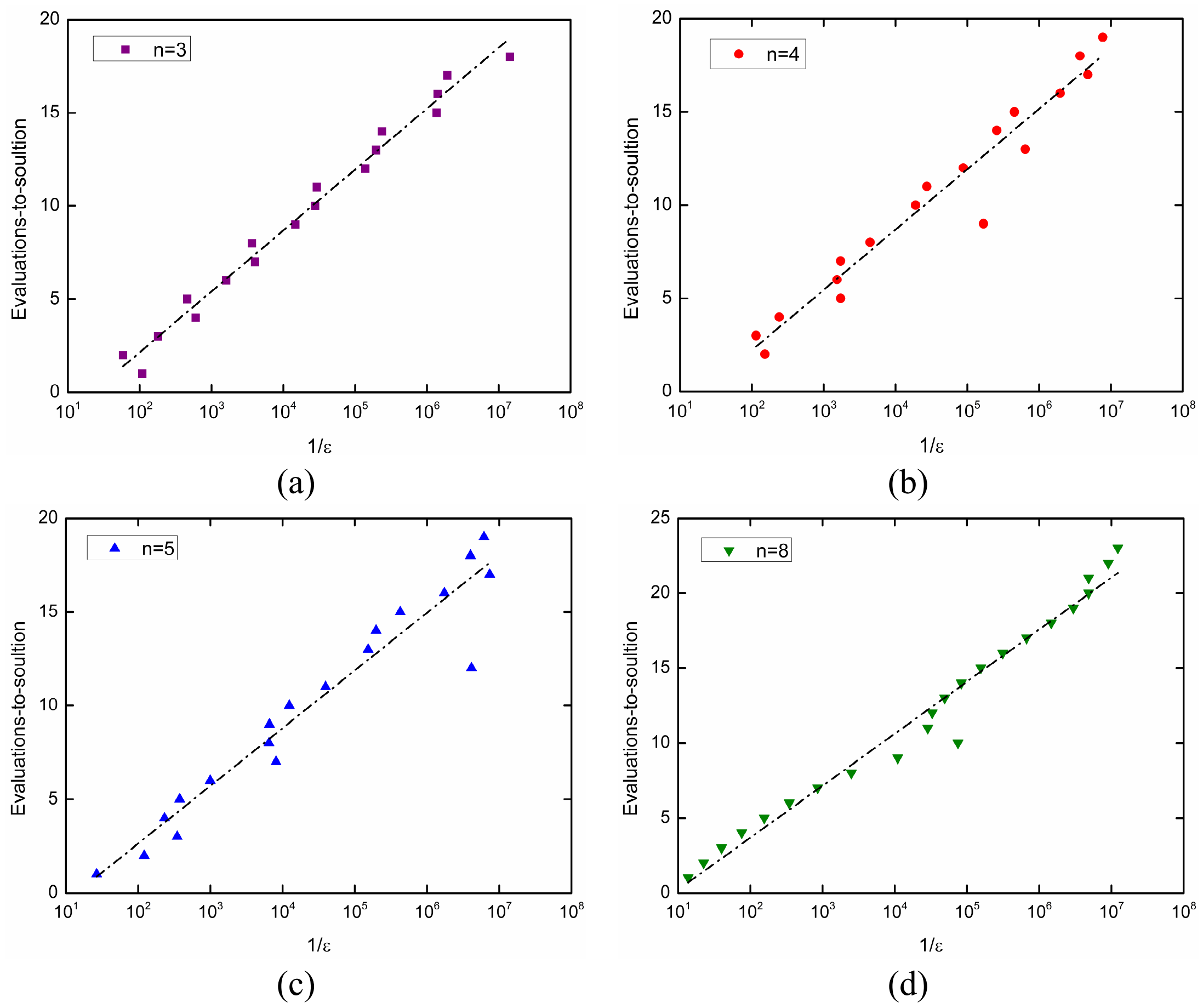

Fig. 10. VQLS heuristic scaling for the matrices generated according to Eq. (22). Evaluations-to-solution versus $1/\varepsilon$ for a system of (a) $n = 3$ qubits, (b) $n = 4$ qubits, (c) $n = 5$ qubits and (d) $n = 8$ qubits. The $x$-axis is shown in a log scale.

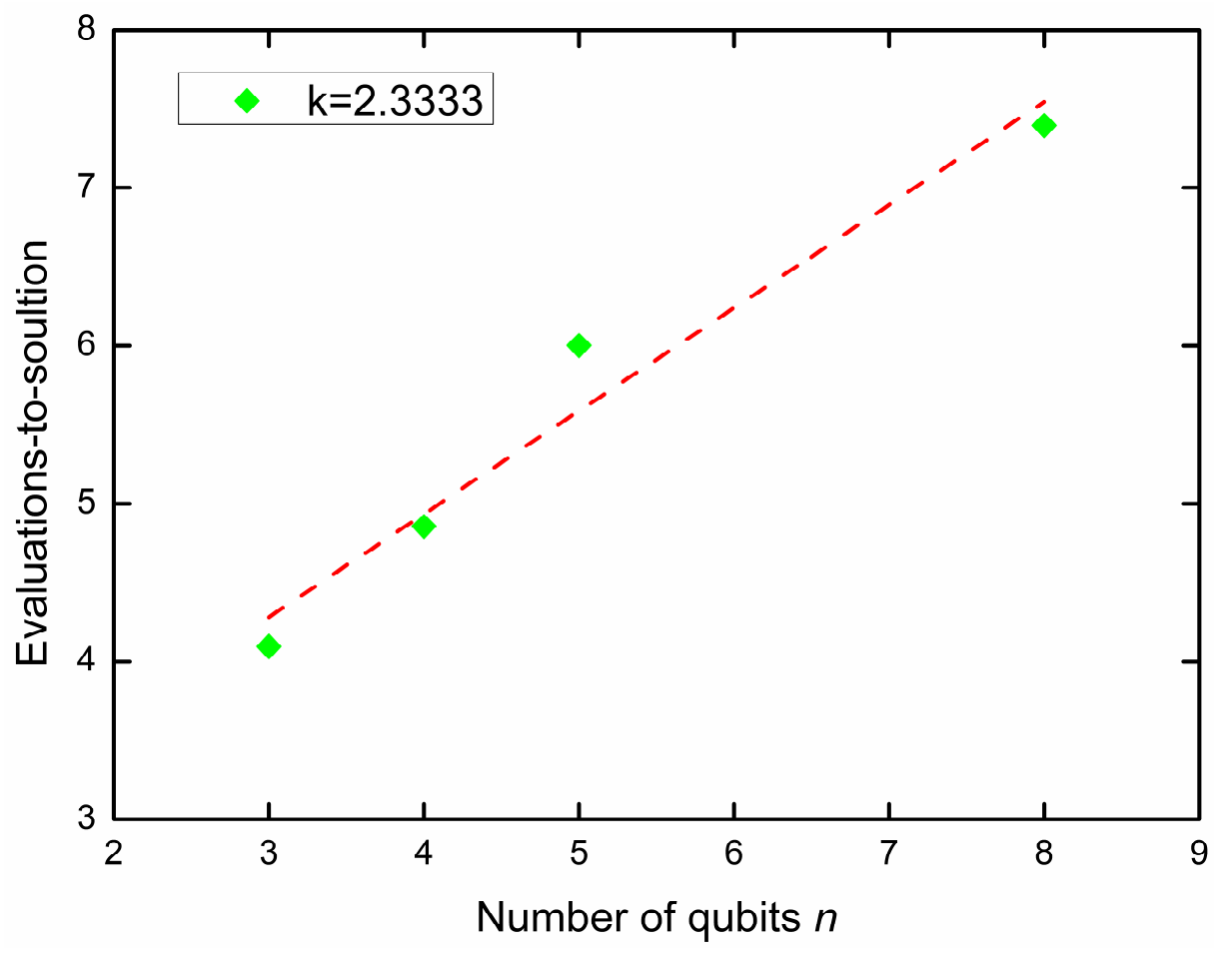

Fig. 11. VQLS heuristic scaling for the matrices generated according to Eq. (22). The evaluations-to-solution is the number of executions needed to guarantee a desired precision $\varepsilon = 0.001$. Evaluations-to-solution versus the number of qubits $n$ for systems of the condition number $k = 2.3333$.

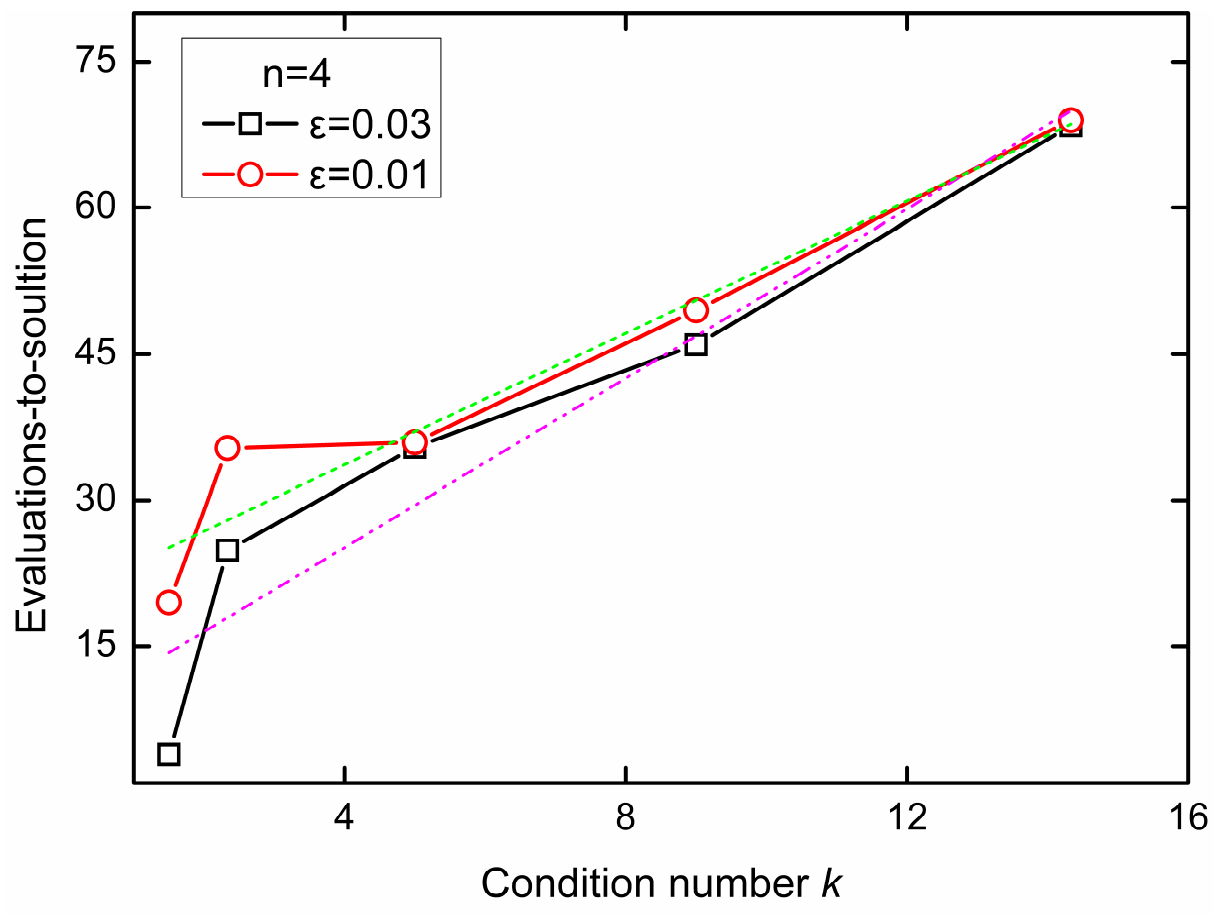

Fig. 12. VQLS heuristic scaling for the matrices generated according to Eq. (22). Evaluations-to-solution versus the condition number $k$ with the number of qubits $n$ = 4.

## 4. Applications to Heat Conduction Problem

The preceding analysis can be extended to simulate the heat conduction problem with one and higher dimensions. Consider the cast linear systems presented in Eqs. (6) and (7). The corresponding implementation details and results with discussions are presented in this section. Note that the discussions hereinafter only include the perfect simulations with the statevector quantum simulator. The analyses of heuristic scaling only consider converged results and each instance is run over 10 times to get the averaged data.

### 4.1. One-dimensional heat conduction

In one-dimensional case, the matrix $A$ shown in Eq. (7) is a strictly tridiagonal matrix. Given the boundary conditions and applying a proper normalization, the quantum states for the matrix $A$ and vector **b** can be obtained. For example, when $N$ = 8 grid points are used to discretize the computational domain, correspondingly $n$ = 3, the matrix $A$ can be linearly decomposed to 8 items as follows.

$$\begin{aligned} A &= \sum_{k=0}^{7} c_k A_k \\ &= 2(I - 0.125X_0X_1X_2 + 0.125X_0Y_1Y_2 - 0.125Y_0X_1Y_2 - \\ &\quad 0.125Y_0Y_1X_2 - 0.25X_1X_2 - 0.25Y_1Y_2 - 0.5X_2). \end{aligned} \tag{25}$$

The state for the vector **b** is prepared by

$$\left|\mathbf{b}\right\rangle = X_0X_1X_2\left|0\right\rangle. \tag{26}$$

The current study is conducted with the number of qubits $n$ = 3, 4, 5 and 8 for the number of interior grid points $N$ = 8, 16, 32 and 256, respectively. For all simulations, the cost function can plateau and the results obtained by the quantum-classical VQLS solver are in acceptable agreement with solutions computed by the classical solver. Fig. 13 shows two simulated results of $N$ = 32 and 256. A slight difference between results of the classical and VQLS solvers exhibits for the case of $N$ = 32. This is within expectation because of the utilization of random initial parameters for the ansatz and the non-optimal optimizer.

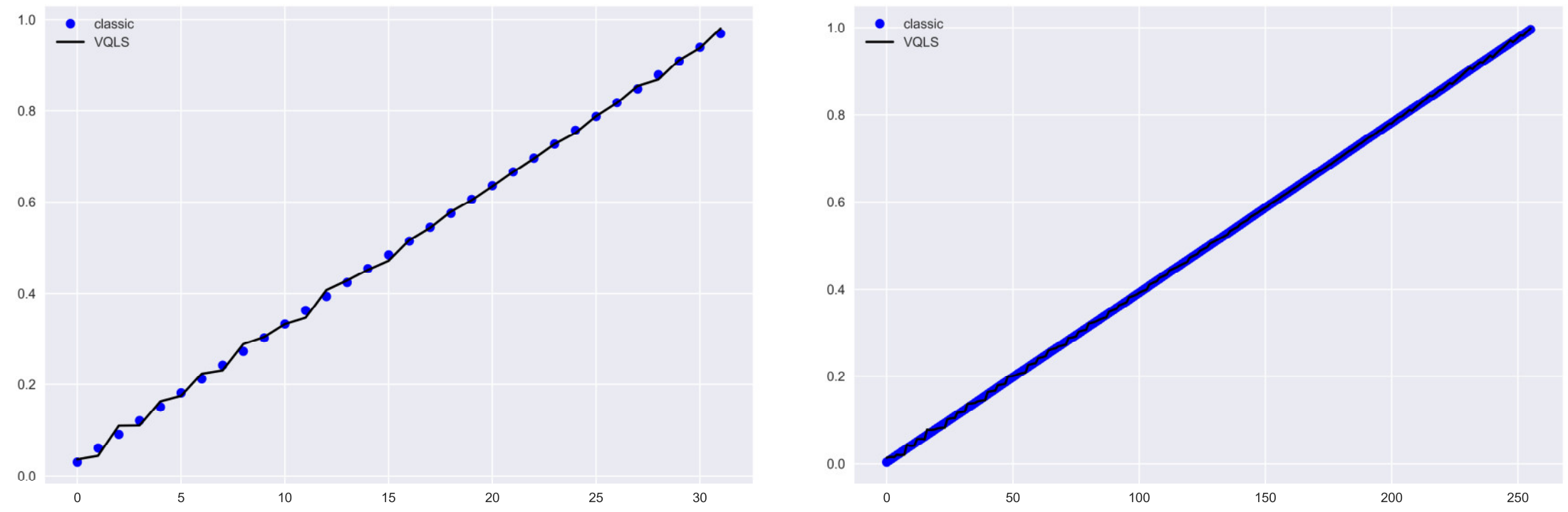

(a) Solutions of temperature when $N = 32$

(b) Solutions of temperature when $N = 256$

Fig. 13. Results of one-dimensional heat conduction problem: solutions of temperature when (a) $N = 32$ and (b) $N = 256$.

Like the analyses in Section 3, the time complexity is further explored for this case via the heuristic scaling. However, since the condition number changes with the size of the matrix (namely $N$) in this problem, we cannot scale the time complexity solely depending on the number of qubits $n$ or the condition number $k$. Hence, we combine their dependences together and mainly focus on the heuristic scaling with $n$ and the precision $\varepsilon$ for this problem.

Through numerical experiments with the similar scaling strategy to that in Subsection 3.2, the relationship for the evaluations-to-solution versus $1/\varepsilon$ is heuristically determined. Furthermore, the scaling with $n$ or $N$ guaranteeing a desired precision $\varepsilon = 0.05$ is found. The corresponding results are presented in Fig. 14 and Fig. 15. Note that the $x$-axis in Fig. 14 is shown in a log scale. As seen from Fig. 14, for all values of $n$, the data nearly can be fitted with a linear function, which implies that the $1/\varepsilon$ scaling is logarithmic. From the heuristic scaling with $n$ for $\varepsilon = 0{:}05$ shown in Fig.

15, the dependence appears to be linear (logarithmic in $N$) for this example.

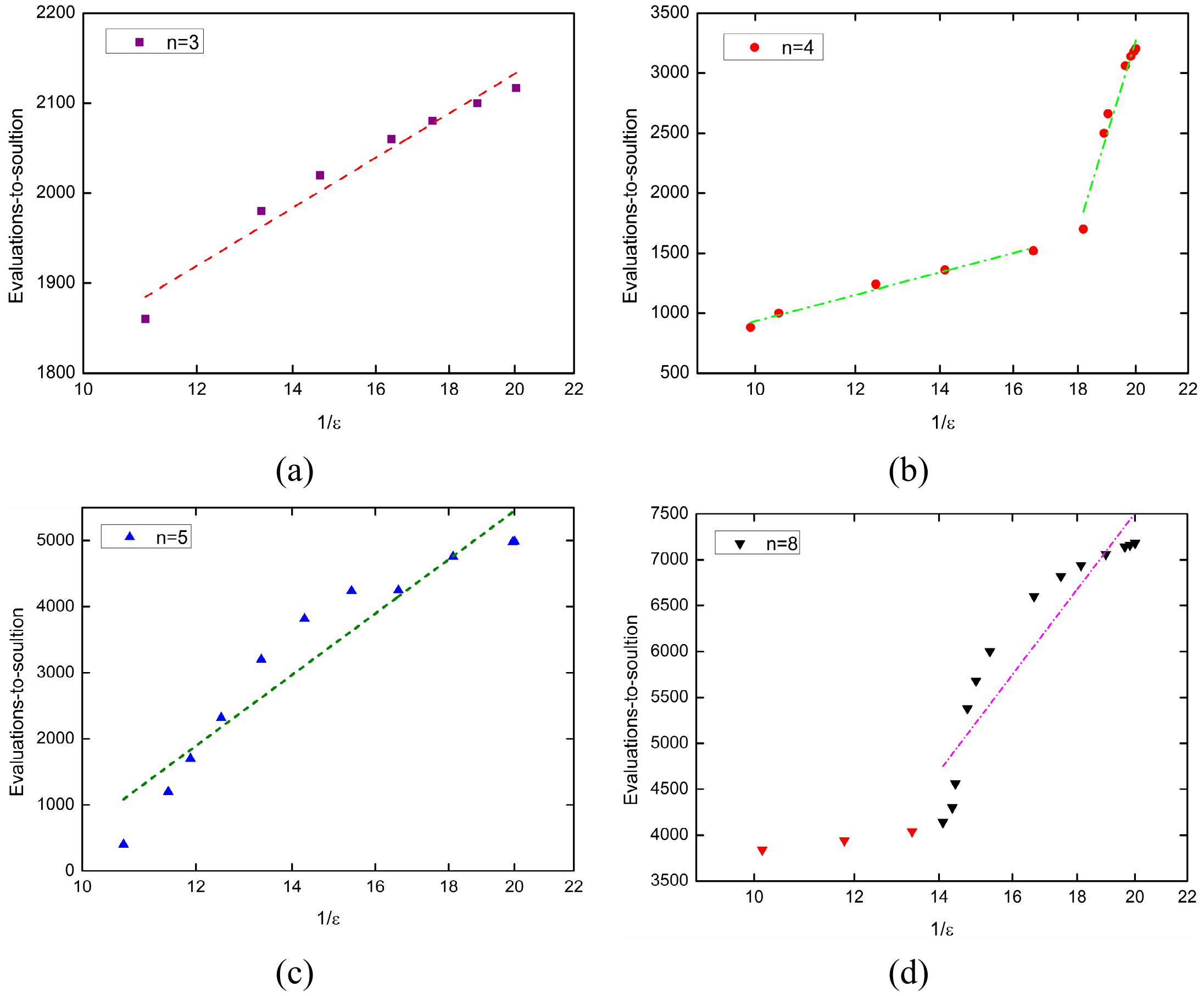

Fig. 14. Heuristic scaling for one-dimensional heat conduction problem. The evaluations-to-solution versus $1/\varepsilon$ for (a) $n = 3$, (b) $n = 4$, (c) $n = 5$ and (d) $n = 8$. The $x$-axis is shown in a log scale.

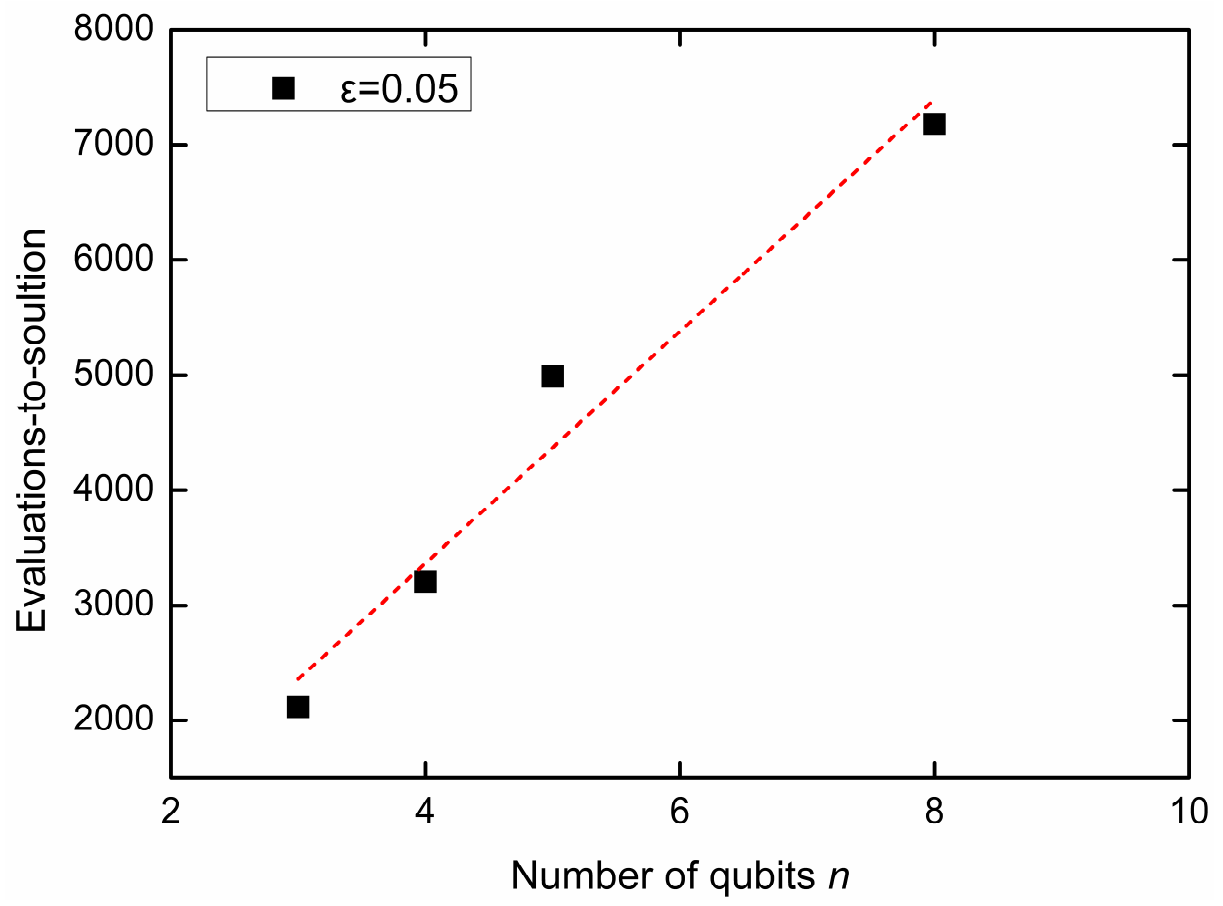

Fig. 15. Heuristic scaling with number of qubits $n$ when $\varepsilon = 0.05$ for one-dimensional heat conduction problem.

### 4.2. Two-dimensional heat conduction

When two dimensions are considered for the heat conduction problem, as shown in Eq. (6), the structure of the sparse matrix $A$ in this linear system becomes a block symmetric banded/Toeplitz matrix. Like the one-dimensional case, the quantum states for the matrix $A$ and vector $\mathbf{b}$ should be prepared via quantum logic operators for execution in quantum circuits. Based on the decomposition method illustrated in Subsection 2.2, when $n = 6$ qubits are used, i.e., $N = 8$ interior grid points in each direction, the matrix $A$ can be linearly decomposed to 15 items. For the case of $n = 8$ and $N = 16$, there are 31 items for the linear combination of unitaries for the matrix $A$.

For this two-dimensional case, three sets of grids with $N = 4$, 8 and 16 corresponding to $n = 4$, 6 and 8, respectively, are used to discretize the computational domain. By successfully applying the quantum-classical VQLS solver, the solutions of temperature as well as the temperature contours are numerically computed. Fig. 16 and Fig. 17 plot the results for cases of $n$ = 6 ($N$ = 8) and $n$ = 8 ($N$ = 16). The reference data computed by a classical solver using the direct method are also included. As shown in the comparisons of the temperature solutions and contours, excellent agreements between the classical and VQLS solvers have been achieved. Such accordance may indicate that the present method based on the VQLS algorithm is competent to solve this heat conduction problem.

Furthermore, the time complexity of the present method is evaluated for this two-dimensional case. Fig. 18 and Fig. 19 depict the heuristic scaling for the dependence on $\varepsilon$ and $n$, respectively. It can be observed that, like the conclusions in the one-dimensional case, the $1/\varepsilon$ scaling is approximately logarithmic and the dependence on $n$ appears to be linear (logarithmic in $N$). This observation provides evident proof of the promising efficiency of the VQLS based method for problems related to CFD.

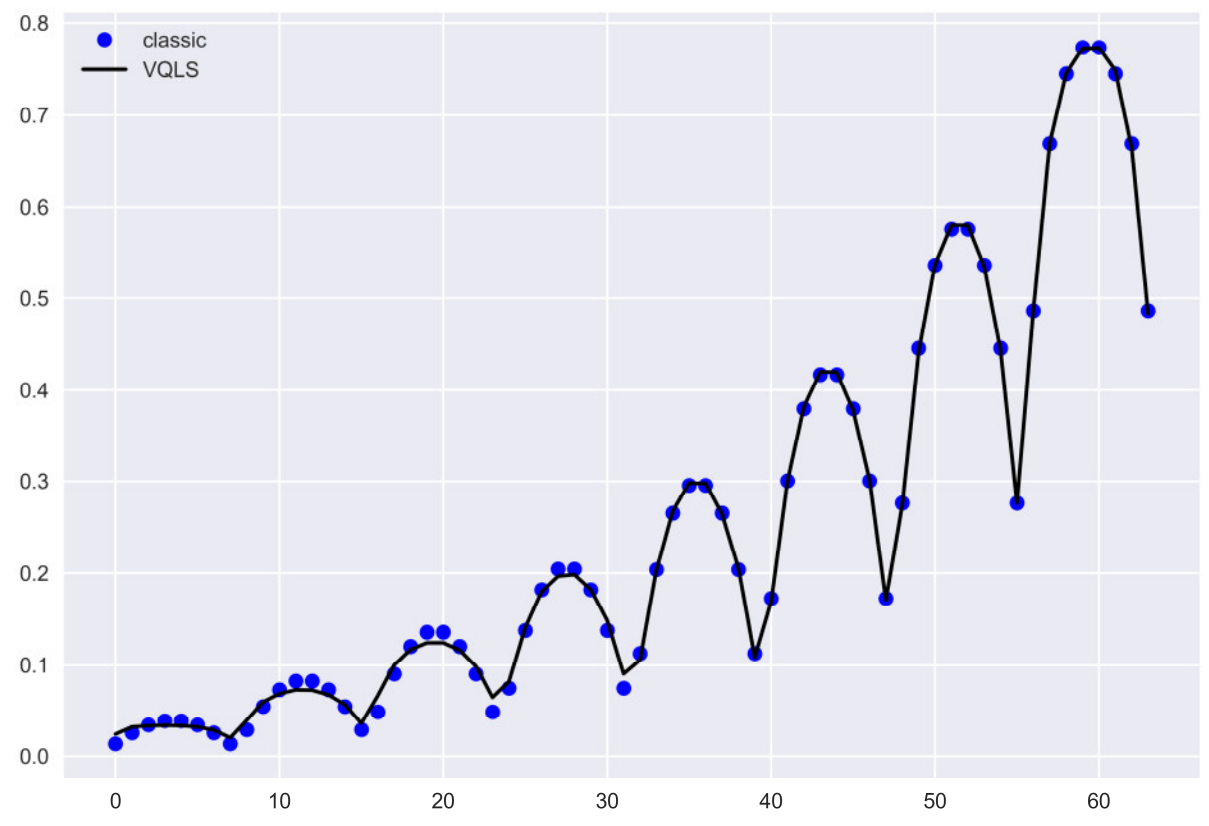

(a)

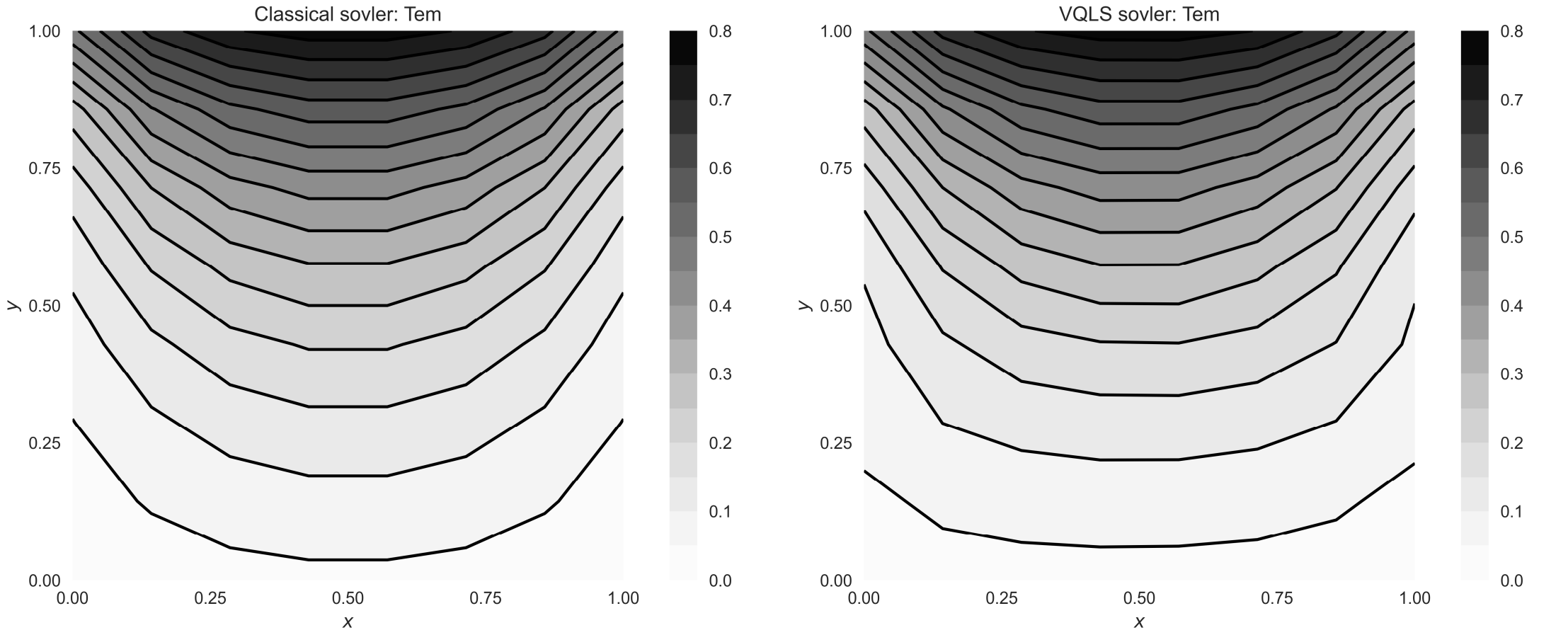

(b) (c)

Fig. 16. Results of two-dimensional heat conduction problem when $N = 8$ and $\varepsilon = 0.05$: (a) comparison of solutions; (b) the temperature contours obtained by the classical solver and (c) the temperature contours obtained by the VQLS solver.

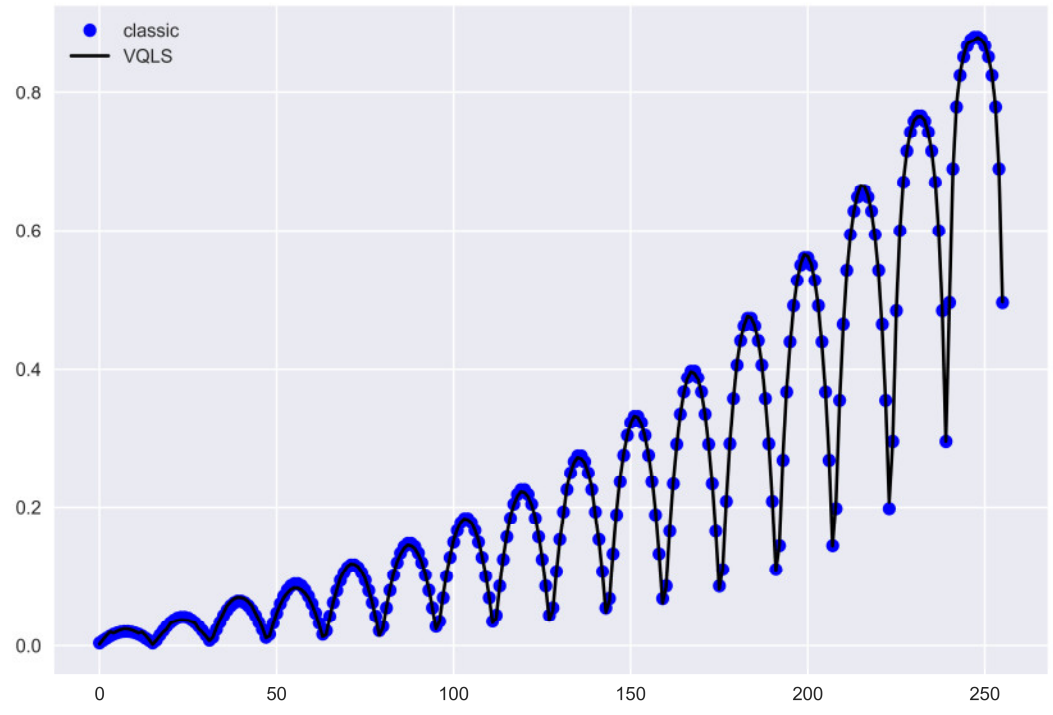

(a)

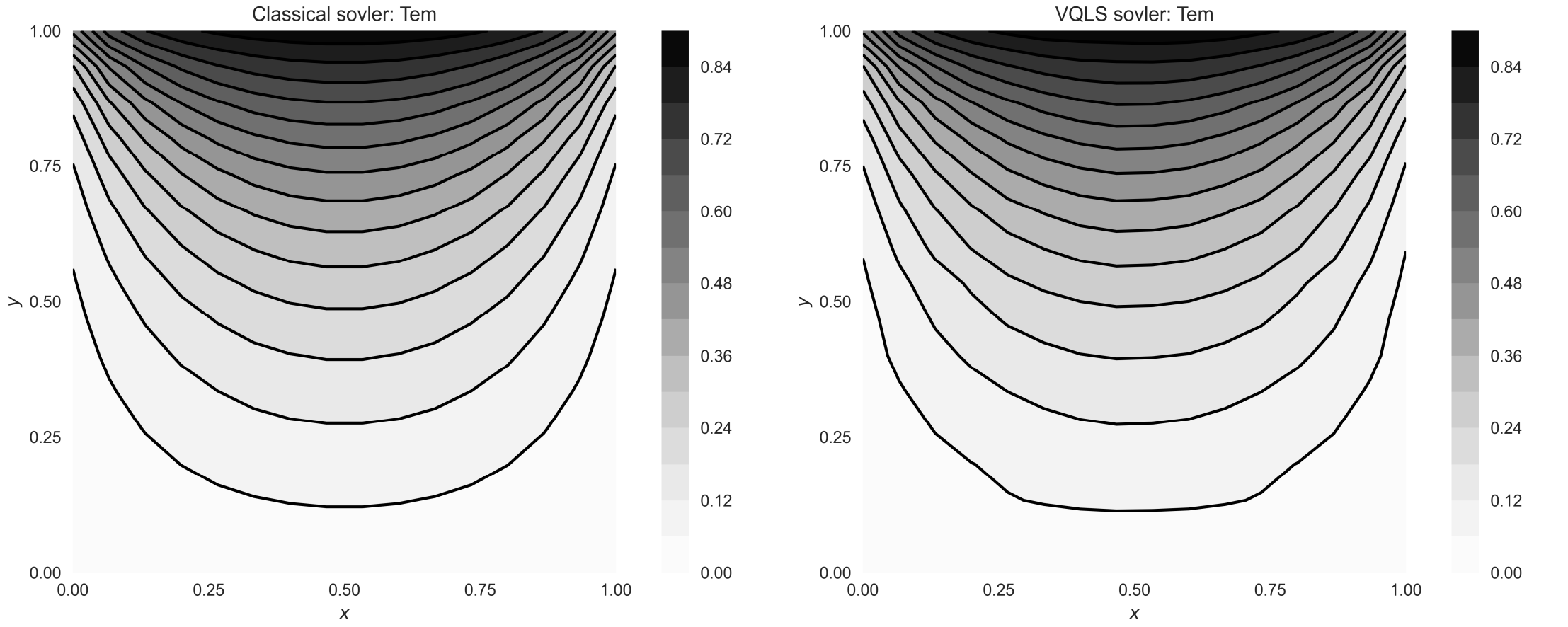

(b) (c)

Fig. 17. Results of two-dimensional heat conduction problem when $N = 16$ and $\varepsilon = 0.05$: (a) comparison of solutions; (b) the temperature contours obtained by the classical solver and (c) the temperature contours obtained by the VQLS solver.

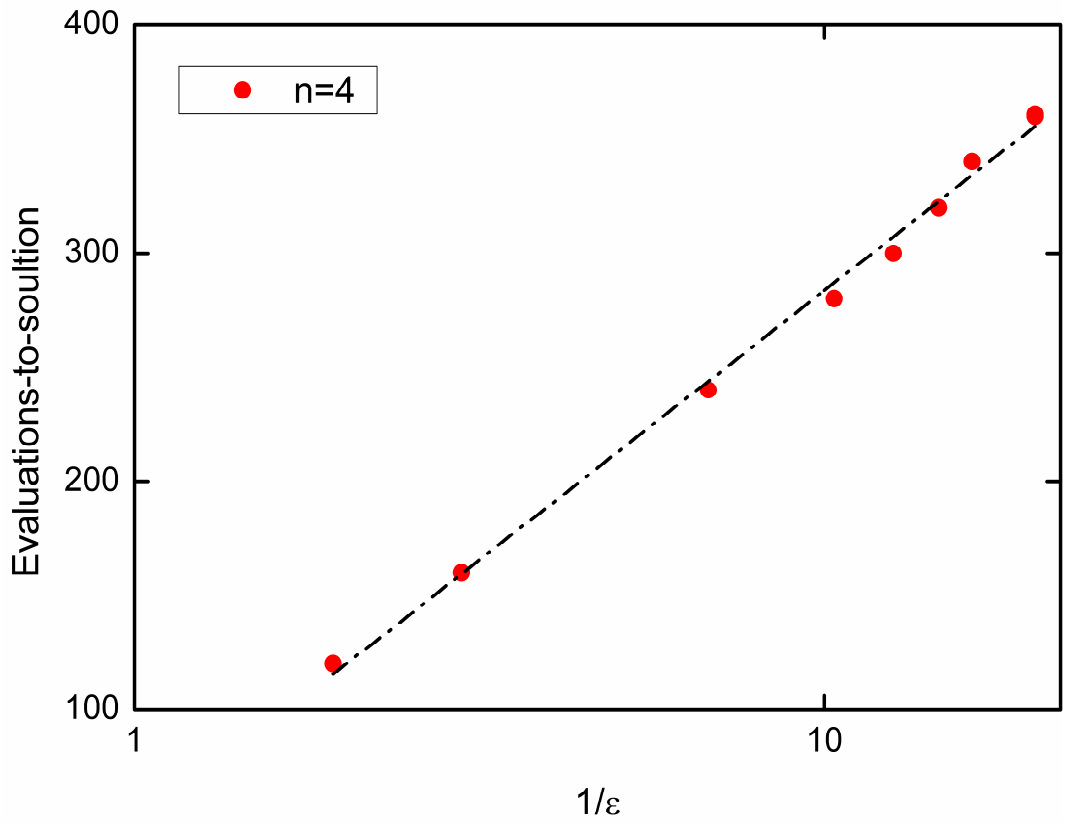

(a)

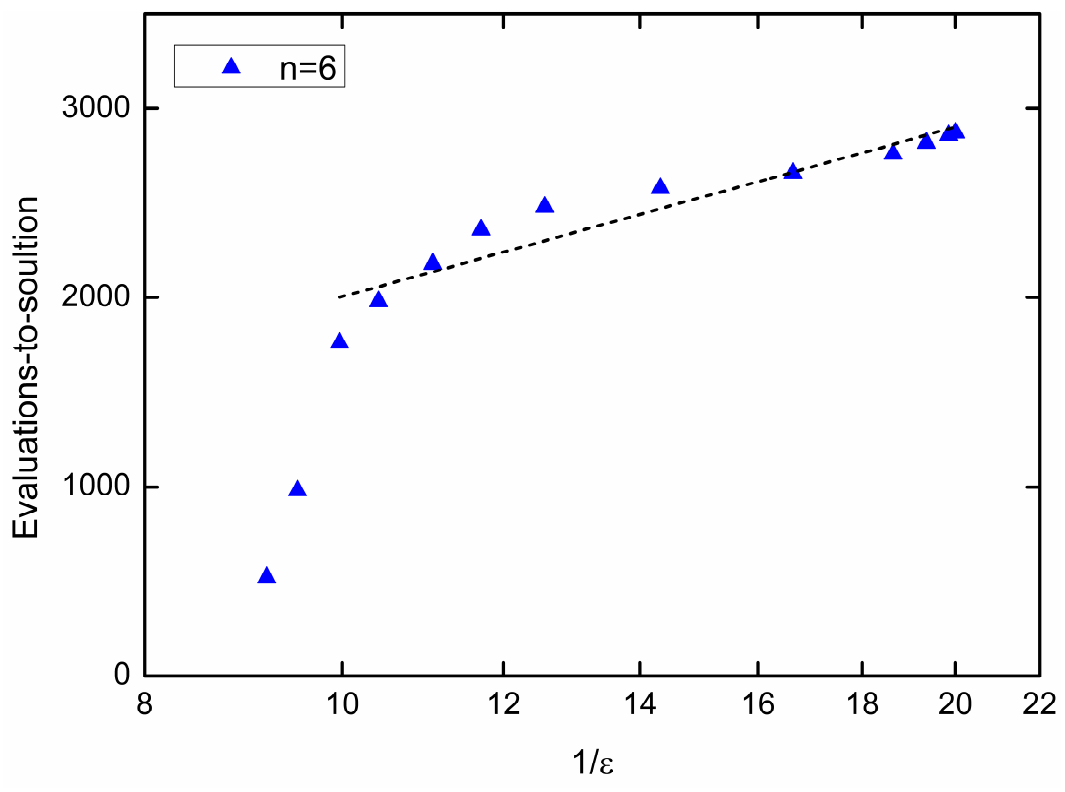

(b)

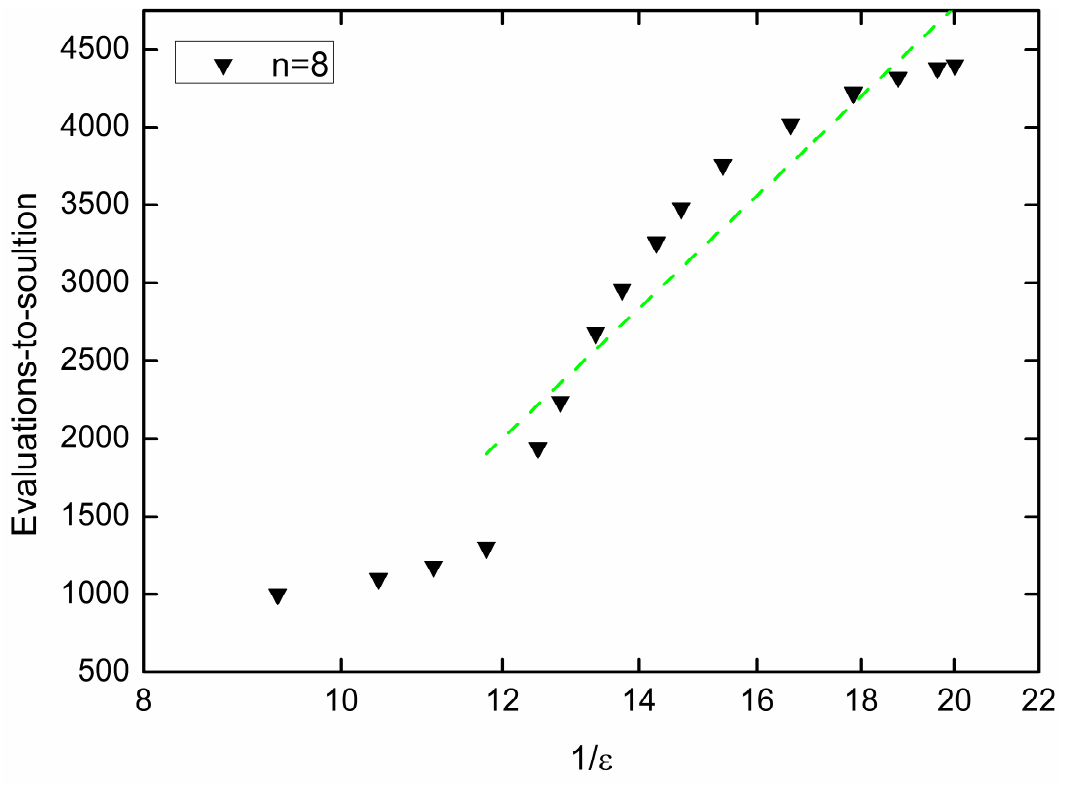

(c)

Fig. 18. Heuristic scaling for two-dimensional heat conduction problem. The evaluations-to-solution versus $1/\varepsilon$ for (a) $n = 4$, (b) $n = 6$ and (c) $n = 8$. The $x$-axis is shown in a log scale.

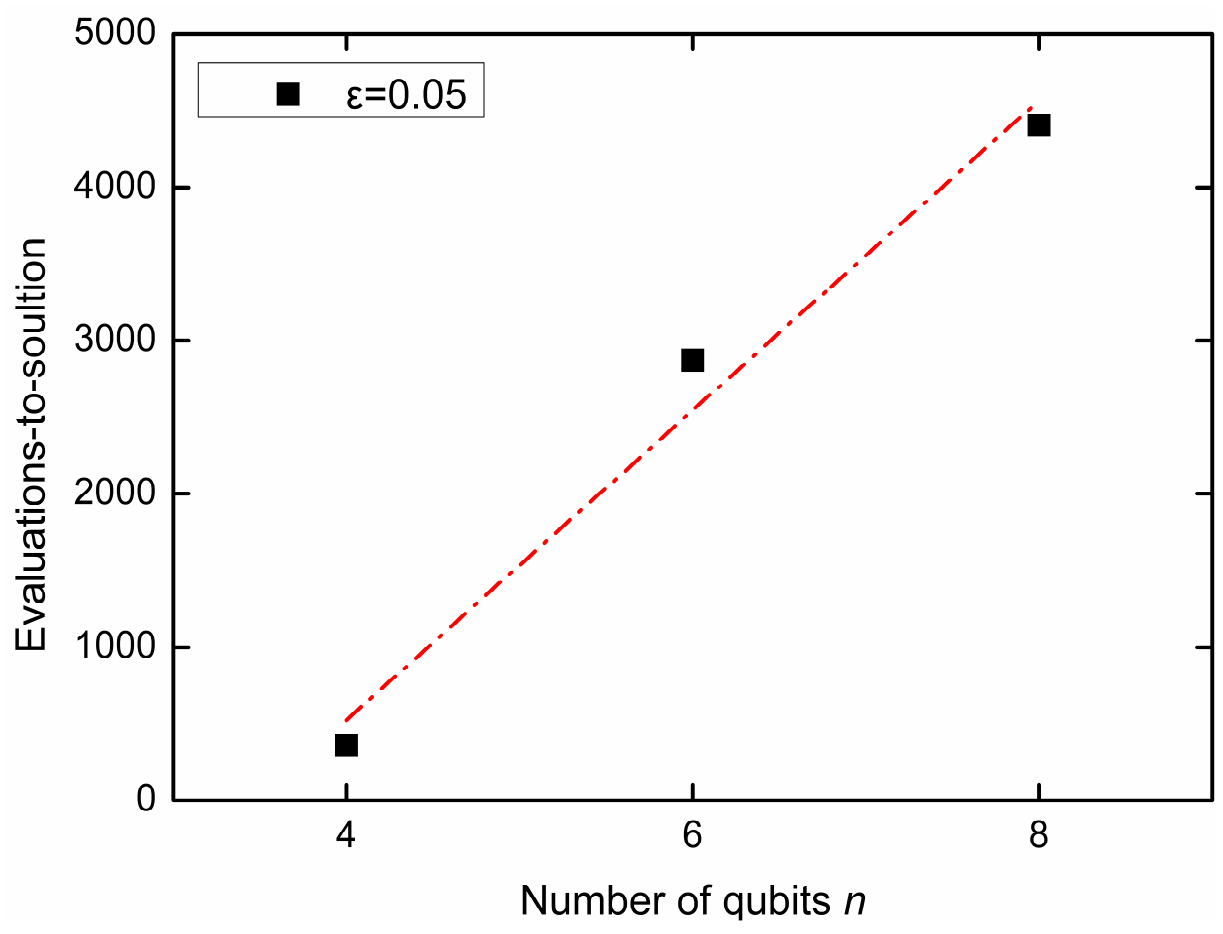

Fig. 19. Heuristic scaling with number of qubits $n$ when $\varepsilon = 0.05$ for two-dimensional heat conduction problem.

## 5. Conclusions and Perspectives

This paper presents a practical application of the variational quantum linear solver for heat conduction equations. Using statevector simulations, it is found that the VQLS algorithm can be useful in numerical resolution of partial differential equations. From the numerical experiments, the relationship between the shots and accuracy is revealed. By doing comprehensive assessment of various parameters, the time complexity of the VQLS algorithm is demonstrated to scale efficiently in the precision, the condition number, and the size of the linear system. In addition, with the successful simulations of the one- and two-dimensional heat conduction problems, the present VQLS based method is well validated in terms of accuracy and efficiency. Based on agreeable results, the heuristic scaling shows the nearly logarithmic dependence on $1/\varepsilon$ and linear dependence on the number of qubits (or logarithmic dependence on the number of interior grids in each direction).

In this study, due to computational limitations, only the relatively small sizes of linear systems are considered. Moreover, the ansatz and optimizer may be non-optimal. Although expected results for the heat conduction problem are obtained by the present quantum-classical solver, it should be admitted that solving linear system of equations in general which results from PDEs is still challenging, and future discussion about the potential speed-up for the restricted and practical tasks would be an interesting research direction.